# MODELLING THE DEEP COUNTS: LUMINOSITY EVOLUTION, DUST AND FAINT GALAXIES

Ana Campos and Tom Shanks
*Physics Dept., Univ. of Durham, South Road, Durham DH1 3LE.*

22 November 1995

**ABSTRACT**

In this paper we analyse the deep number counts problem, taking account of new observational and theoretical developments. First we show that the new Bruzual and Charlot(1993) models allow a new class of spiral dominated luminosity evolution (LE) model where significant amounts of the luminosity evolution needed to fit faint count data are due to spiral rather than early-type galaxies. Second we show that the inclusion of dust may be a vital ingredient for obtaining fits with any LE model. Third we compare the quality of fit of both the spiral and early-type LE models, including dust, for a wide variety of observational data. We find that parameters can be found for both LE models which allow a good fit to all data with the exception of the faintest $B > 25$ counts in the case of $q_o = 0.5$ cosmologies, where some luminosity dependent evolution may be needed (see also Metcalfe et al 1995). Otherwise both these classes of LE model, with the inclusion of dust, provide an excellent foundation for understanding the $B < 25$ galaxy counts and galaxy counts and redshift distributions in a variety of other wavebands.

## 1 INTRODUCTION

Counting galaxies is one of the widely-used observational methods in cosmology to constrain the evolution of galaxies and the value of the deceleration parameter $q_0$ (throughout this paper we will assume $\Lambda = 0$). The observed number of galaxies per interval of magnitude and unit area is compared with the predictions from different cosmological models. In the simplest case, a basic assumption is made: namely that the luminosity and the number density of galaxies remain both constant with the time. Then, for a given luminosity function (LF) estimated using nearby galaxy redshift surveys is straightforward to compute the number count predictions by accounting for the redshift-dependence of the volume element in the different geometries considered. These basic models are usually called non-evolving.

Since the early works on number counts it was realized that non-evolving models failed to fit the counts, no matter the value of $q_0$ considered (eg. Shanks et al. 1984; Koo 1986; Tyson 1988). There is an *excess* of galaxies at faint levels with respect to the model predictions, which is larger when the counts are made using blue photometric bands.

A plausible way to overcome the discrepancy between observations and model predictions is by allowing evolution in the galaxy population (eg. Tinsley 1980). The luminosity of galaxies is expected to evolve with time (Guiderdoni & Rocca-Volmerange 1990; Metcalfe et al. 1991), and the evolution follows different paths depending on the morphological type. The number of galaxies per unit volume could also vary (Guiderdoni & Rocca-Volmerange 1991; Broadhurst et al. 1992), like for example in hierarchical models of galaxy formation where galaxy sub-units merge to build up the present day population.

In this paper we re-visit the problem of the deep counts by analyzing all different ingredients which must be taken into account to model the number counts. We take advantage of the latest observational developments (eg, deep galaxy redshift surveys; Glazebrook et al. 1995; Campos et al. 1995; Crampton et al 1995) which can be more than useful to interpret the faint counts. In §2 we discuss the so-called luminosity evolution and merging models. The role of dust in modelling the galaxy luminosity evolution is analyzed in §3. In particular we show how the disagreements between the luminosity evolution models and the observations are smoothed out when internal extinction (dust) in galaxies is considered. In §4 we find that the $q_0 = 0.5$ model even with the inclusion of dust underpredicts the B> 25 counts. The inclusion of luminosity dependent evolution of the form discussed by Metcalfe et al. (1995) is still required to fit the faint counts. Finally we devote §5 for an overview of the models presented in the previous sections.

## 2 DEEP NUMBER COUNTS: A HINT FOR GALAXY EVOLUTION

### 2.1 Luminosity evolution and merging models

The steep slope of the counts at its faint end was first interpreted as a hint that the luminosity of galaxies is evolving with time. As galaxies were brighter in the past, for a given *apparent* magnitude range they are observable at redshifts higher than in the non-evolving case. Then *Pure* Luminosity Evolution (PLE) models predict larger numbers of galaxies to be observed at faint magnitudes, eg. Tinsley 1980; Bruzual & Kron 1980; Koo 1981; Shanks et al. 1984; Wyse 1985; King & Ellis 1985; Yoshii & Takahara 1988; Guiderdoni & Rocca-Volmerange 1990; Metcalfe et al. 1991, 1995. (Although based upon different spectrophotometric models of stellar population - basically Tinsley 1972; Bruzual 1981, 1983; Arimoto & Yoshii 1986, 1987 - most of these works assume an exponentially decaying star formation rate SFR for the early type galaxies and a nearly constant SFR for the late ones). The evolution of the luminosity is much larger as we observe in bluer photometric bands, and so the difference between the predictions from PLE and non-evolving models is larger in the blue than in the near-IR.

The main criticism of PLE models came from the fact that they predict a high mean redshift for the faint galaxies. The available redshift surveys (Broadhurst et al. 1988; Colless et al. 1990, 1993; Glazebrook et al. 1994; Campos et al. 1995) show a deficit of high-redshift galaxies compared to the predictions from PLE models.

In order to increase the number of galaxies seen at faint limits without increasing the average redshifts of the objects, i.e. to be consistent with the number counts $N(m)$ and the redshift distributions of faint field galaxies $N(z)$ simultaneously, Guiderdoni & Rocca-Volmerange (1991) and Broadhurst et al. (1988; 1992) proposed the so-called merging models. In these models both the luminosity and the number density of galaxies evolve with time, and so galaxies are built up by subsequent mergers of sub-units. As an example, in the model proposed by Broadhurst et al. a present day galaxy is the result of the merging of $\sim 4$ sub-units that existed at $z \sim 1$. Then, in this model the *excess* of faint field counts is just due to the change of $\phi_*$, which decreases with time. Glazebrook et al. (1995) argued that a major success of the merging model is the fit provided to $N(z)$ of $B : 23 - 24$ galaxies. The observed distribution has a shape that resembles the predictions from non-evolving models, although with a much higher normalization. As in the merging models the *excess* of blue counts is explained in terms of $\phi_*$ evolution, the $N(z)$ prediction has similar shape than non-evolving but with higher normalization.

Campos et al. (1995) have pointed out that the observed $N(z)$ of $B : 23-24$ galaxies should be approached cautiously, as the redshifts of many ($\sim 40\%$) objects could not be identified. As we will later comment in more detail, in this paper it is argued that many of the unidentified galaxies might be placed at high-z, and so the *observed* $N(z)$ distribution could be different to the *real* one. This view is supported by the work of Roche et al. (1993; 1995), who studied the scaling of the amplitude of the angular correlation function $\omega(\theta)$ with depth, finding that it falls well below the predictions from non-evolving models, or in general from any model giving a

**Table 1.** Galaxy Models and Luminosity Functions

| Type | $\phi_*$ (%) | $\alpha$ | $M_*$ |
|---|---|---|---|
| E/S0/Sa | 57% | -0.7 | -19.6 |
| Sb/Sc | 26% | -1.1 | -19.9 |
| Sd/Irr | 17% | -1.5 | -20.0 |

The Luminosity functions for the three types of galaxies.

$N(z)$ shape similar to the non-evolving one, as is the case of merging models.

Besides, and as already noticed by Guiderdoni & Rocca-Volmerange (1991), the *price to be paid* in fitting the counts with merging models is a strong number density evolution. Such a large amount of merging might be in disagreement with the properties of present-day galaxies, in particular the stability of spiral disks.

In what follows we further analyze in detail PLE models. First we will show that they fail in fitting $N(z)$ but also the deep faint counts. However, as illustrated in §3, when the extinction by dust is considered PLE models are able to provide excellent fits to both $N(m)$ and $N(z)$ without any need for number density evolution. We only find neccesary to include either merging or the presence of a large population of dwarf galaxies to fit the counts at B> 25 if $q_0 = 0.5$.

### 2.2 New PLE models

We have computed predictions from PLE models for the number counts and redshift distributions using the new set of spectrophotometric models by Bruzual & Charlot (1993). Following Metcalfe et al. (1991), galaxies were divided into three main types: early, intermediate and late corresponding to E/S0/Sa, Sb/Sc and Sd/Irr respectively (we shall refer to them as E, S and I). For each type we take a different luminosity function obtained from the Durham-z survey (Shanks 1990; Metcalfe et al. 1991; See Table 1). We consider two different sets of models for the SFR , which is taken to decay exponentially with a different e-folding time for each type. For the first set of models (E-model) we take $\tau_E = 1.2$, $\tau_S = 900$ and $\tau_I = 900$ Gyr, whereas for the second one (S-model) $\tau_E = 0.5$, $\tau_S = 7$ and $\tau_I = 900$ Gyr. The age of galaxies at present is assumed to be 16 Gyr and 13/12.4 Gyr for the $q_0 = 0.05$ and 0.5 models respectively. For $H_0$=50 km s$^{-1}$ Mpc$^{-1}$ this corresponds to a redshift of formation of $z_{for} = 6$ for $q_0 = 0.05$ and $z_{for} = 35/6$ for $q_0 = 0.5$. The Initial Mass Function (IMF) is the Salpeter one in all the three cases and at any time.

The reason to choose these two sets of SFR is the following. The $K+e$ corrections for model-galaxies in the E-model are very close to those used in Metcalfe et al. (1991; 1995), who were able to provide excellent fits to both the number counts and redshift distribution of faint galaxies. In this model the excess of counts over the predictions from non-evolving models is explained in terms of (strong) luminosity evolution of elliptical galaxies. On the other hand, Bruzual & Charlot recommend the use of the $\tau = 7$ Gyr model for intermediate-type galaxies with a single Salpeter IMF. The $\tau = 7$ Gyr model has more B-band evolution than the previous Bruzual (1981) ones. This offers up the new possibility of obtaining significant evolution from the spirals.

In Figure 1 we show the $K + e$ (i.e. K-correction plus evolution) corrections. As already mentioned, notice that for the E-model (i.e. $\tau_E = 1.2$, $\tau_S = 900$ and $\tau_I = 900$ Gyr) the $K+e$ corrections are very close to those used by Metcalfe et al. (1991; 1995), which were computed from the old Bruzual models (Bruzual 1981). The reason to re-compute the predictions for the number counts using the new set of models is the availability of detailed tracks to the redshift of galaxy formation.

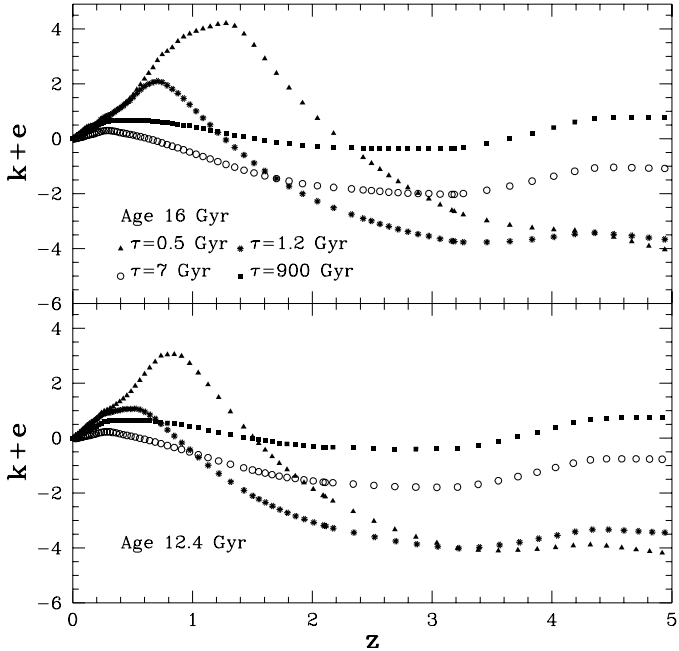

**Figure 1.** K-correction plus evolution for four types of model-galaxies. The age of the galaxies at $z = 0$ is 16 (top panel) and 12.4 Gyr (bottom panel). This corresponds to a redshift of formation of $z = 6$ for $q_0 = 0.05$ and $q_0 = 0.5$ respectively. $H_0$ is assumed to be 50 km s$^{-1}$ Mpc$^{-1}$.

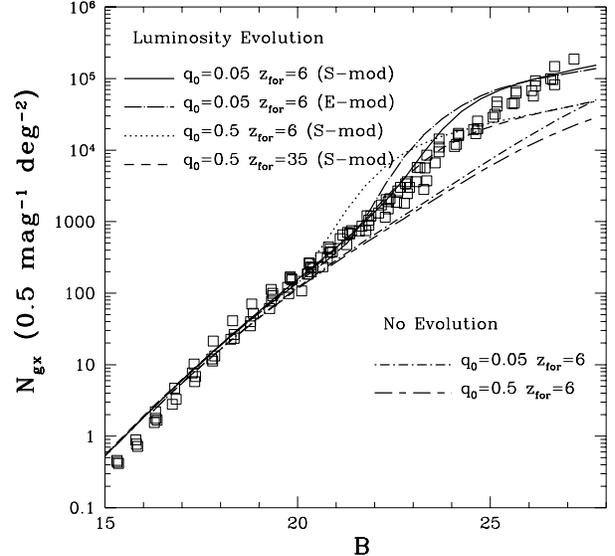

**Figure 2.** The number counts in the B band. The data has been taken from the literature (see text for references). Also we plot predictions from non-evolving models and PLE models.

### 2.3 Comparison of New PLE models and Data

The observed number counts in the B-band are shown in Figure 2. The data is a compilation from the literature: (Kron 1978; Koo 1986; Jarvis & Tyson 1981; Couch & Newell 1984; Infante et al. 1986; Tyson 1988; Jones et al. 1990; Lilly et al. 1991; Metcalfe et al. 1991, 1995). In the same Figure we also show predictions from non-evolving (for comparison) and PLE models. The non-evolving models have been computed using the K-correction polynomial coefficients as given in Metcalfe et al. (1991). All models have been normalized to fit the counts at $B = 17.5$, which corresponds to an overall normalization of $\phi^* = 1.4 \times 10^{-2}$ Mpc$^{-3}$.

None of the models provides a reasonable fit to the counts. As already known, the non-evolving models under-predict the counts. With respect to the PLE models, the $q_0 = 0.5$ models under-predict the counts at faint limits, whereas the $q_0 = 0.5$ with $z_{for} = 6$ and the $q_0 = 0.05$ ones show a *bump* at faint magnitudes which is not seen in the data. The *bump* is shifted from one model to the other. This is due to the differences in geometry from model to model but also because the age and the SFR sets differ, and so the $K + e$ corrections (see Figure 1).

It is important to notice that the predictions shown in Figure 2 from the $q_0 = 0.05$ E-model are different to those shown in Metcalfe et al. (1991; 1995) even if the $K+e$ corrections are very similar for the three galaxy types, as we said before. The reason is found in the $K + e$ corrections beyond $z = 2$. As the old spectrophotometric models (Bruzual 1981) were only tabulated in the range $0 < z < 2$, the $K+e$ correction was assumed to remain constant beyond it. However, as can be seen in Figure 1, $K+e$ changes beyond $z = 2$, particularly for the early type galaxies. As we approach the redshift of formation, the early galaxies for which the e-folding time is assumed to be quite small become much brighter in a relatively short period of time. In fact the responsible for the predicted *bump* in the number counts are the evolved high-redshift early type galaxies, many of them beyond $z = 2$. We will go back to this point in the next section. The same problem (over-prediction of counts at $B \sim 23 - 25$) is found for the S-model as well, although now the *bump* is due to both the early and intermediate type galaxies at high redshift.

The predictions from both PLE models for the redshift distribution of faint galaxies are also unable to fit the data, as they predict a very large *high redshift tail* which is not observed. As an example, the number of galaxies with apparent magnitude $23 < B < 24$ predicted to be at $z > 1$ is $\sim 70\%$, in contradiction with the observations (Glazebrook et al. 1995; Campos et al. 1995).

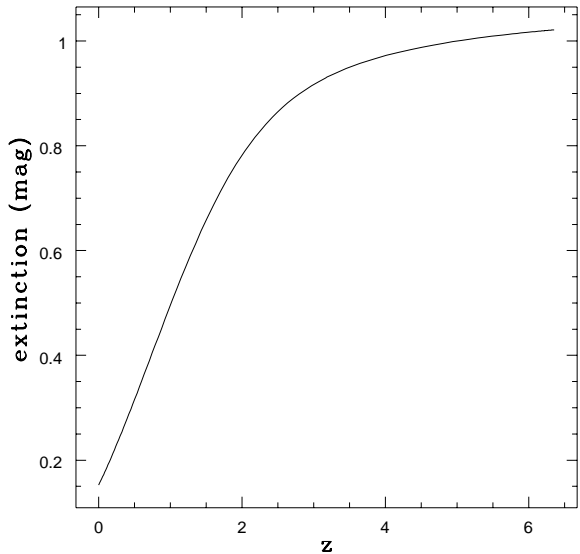

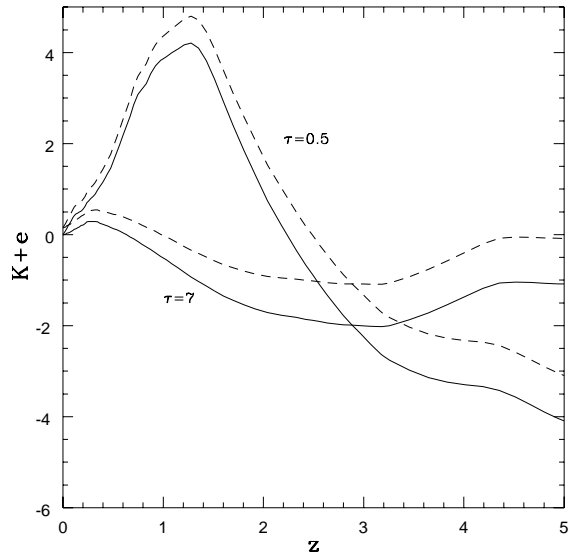

**Figure 3.** Extinction in the B-band as a function of the redshift, for an $L^*$ galaxy.

**Figure 4.** The unreddened (solid lines) and reddened (dashed lines) $K+e$ corrections for an $L^*$ galaxy, for two different models: $\tau = 0.5$ and $\tau = 7$ Gyr.

## 3 THE ROLE OF DUST IN NUMBER COUNTS PREDICTIONS

### 3.1 Dust, B-band Counts and Faint Galaxies Redshifts

The effect of absorption by interstellar dust within galaxies has been first explored by Wang (1991). The presence of dust is also considered by Gronwall & Koo (1995) in their luminosity evolution models. Next we study in some detail how the introduction of a dust component in the luminosity evolution of the galaxies modifies the model predictions. To model the effect of dust we consider an optical depth of $\tau^* = 0.3$ for present-day $L^*$ galaxies (somewhat higher than Wang 1991), and the standard $\lambda^{-2}$ extinction law (Draine & Lee 1984). We also assume that the optical depth depends on the galaxy luminosity as $\tau \propto L_{z=0}^{0.5}$. In Figure 3 we plot the amount of extinction (in magnitudes) in the B-band for an $L^*$ galaxy as a function of the redshift, and in Figure 4 the unreddened and reddened $K + e$ corrections for two model-galaxy types, $\tau = 0.5$ and 7 Gyr, for an $L^*$ galaxy as well. Notice that even if the amount of dust is considered constant with time (i.e. we do not include dust evolution), the extinction depends on the redshift due to the wavelength dependence of the extinction law.

In Figure 5 we plot the B number counts together with predictions from the PLE models shown in Table 1, but taking into account the presence of dust (PLE+D in what follows). As can be seen, the fits are now better as the *bumps* present in the PLE models have vanished.

The dust absorbs the light coming from the stars, re-radiating it in the infrared. The absorption is wavelength selective, becoming more important at shorter wavelengths. This means that the effect of dust is much more severe the higher is the redshift of the galaxy, because of the redshifting of the spectral light. Then, even *small* amounts of dust within galaxies can have very important effects to predict the number counts, because they reduce the luminosity evolution since the brightening is compensated by the extinction.

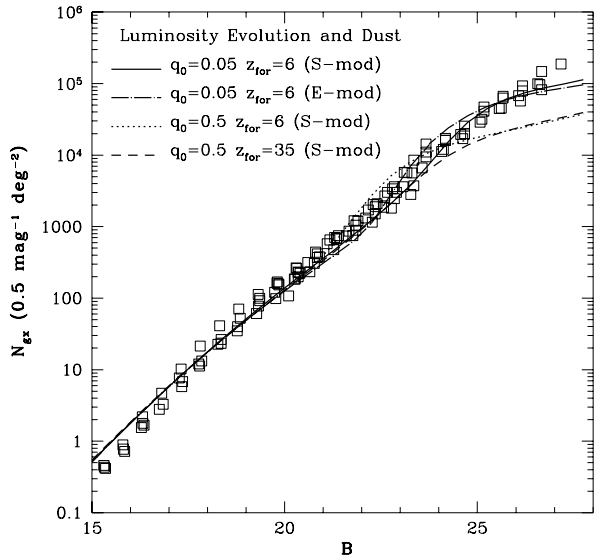

**Figure 5.** The B-band number counts and predictions from PLE+D models.

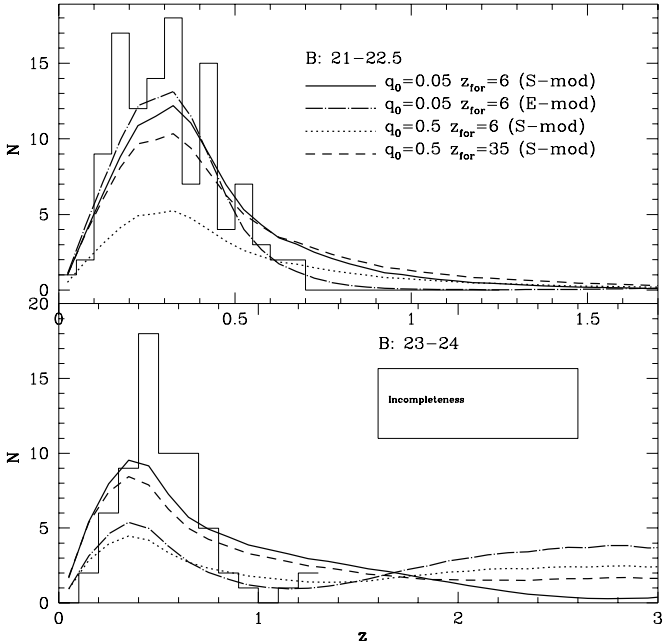

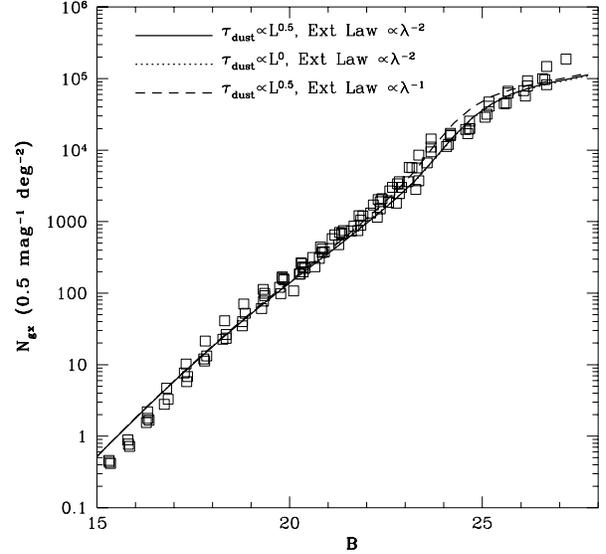

Figure 6. The redshift distribution of faint field galaxies selected in the B-band, and predictions from the models shown in Figure 5. The models have been normalized to the total number of objects in the sample (i.e. identified plus unidentified galaxies)

Figure 8. The B-band number counts and predictions from different $q_0 = 0.05$ - $z_{for} = 6$ PLE+D S-models

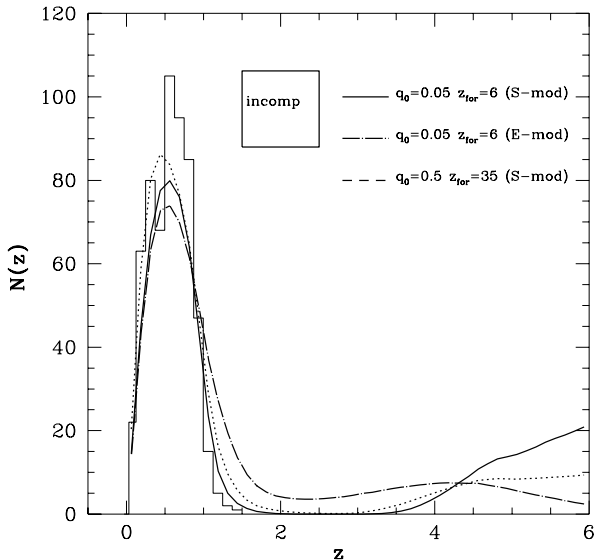

Figure 7. The same as Figure 6, but for galaxies selected in the I-band.

The effect is further observed in the $N(z)$ predictions, plotted in Figures 6 and 7 for galaxies in the magnitude ranges $b_J = 21 - 22.5$ $B : 23 - 24$ and $I : 17.5 - 22.5$ respectively. In the $B : 23 - 24$ range the models predict a *high redshift tail,* but now the percentage of galaxies expected to be seen at $z \gtrsim 1$ is smaller, of the same order of the incompleteness rate of the sample. In fact, the model predictions would nicely fit the data if the unidentified galaxies in the $B : 23 - 24$ sample were located at high redshift, as suggested by Campos et al. (1995). This view is in agreement with the recent work by Cowie et al. (1995), who have found a substantial fraction of star-forming $B \sim 24$ galaxies at $z \sim 1 - 2$. The fit to the $N(z)$ distribution of I-selected galaxies is also reasonable, this survey having the advantage that the incompleteness rate is much smaller ($\sim 15\%$).

Both the $q_0 = 0.05$ E-model and S-model provide reasonable fits to $N(z)$ and $N(m)$. The E-model over-predicts the number of high-z galaxies in the $B : 23 - 24$ sample, as can be seen in Figure 6. However the simplicity of the evolutionary model, in particular the dust model, makes it difficult to rule out the E-model on the basis of these discrepancies. As an example, if the amount of dust evolved with time in early type galaxies, such that they were slightly more *obscured* at high-z, then the predicted high-redshift tail would be reduced, in accordance with the data. With respect to the $q_0 = 0.5$ model, we show in Figures 5 and 6 predictions from the S-model for two different galaxy formation epochs:

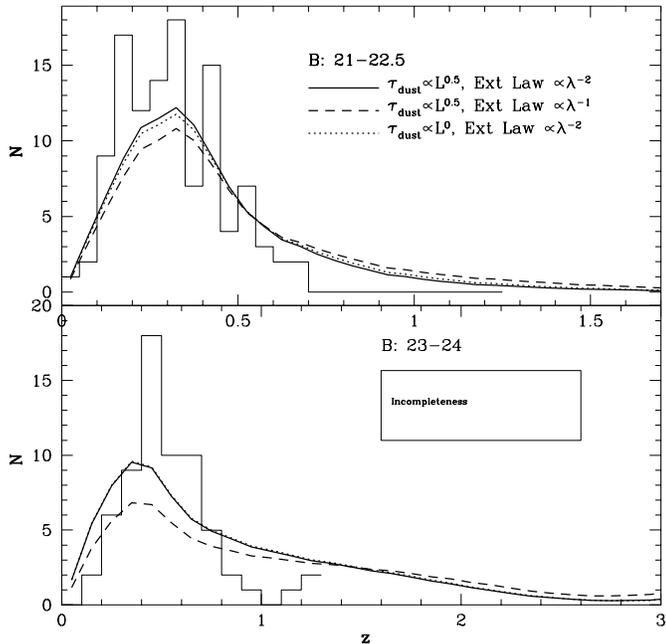

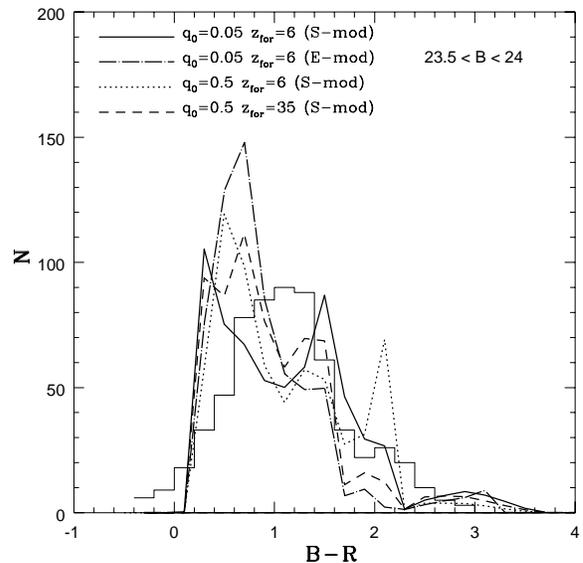

**Figure 9.** The redshift distribution of B-selected faint galaxies, and predictions from the models shown in Figure 8

**Figure 10.** B−R colour distribution of faint galaxies and predictions from the PLE+D models shown in Figure 6.

$z_{for} = 6$ and 35. The $z_{for} = 35$ model provides better fits to both $N(m)$ and $N(z)$. However, as we will show later, by changing slightly the SFR for elliptical galaxies it is possible to provide better fits for lower formation redshifts. In fact, as the Universe is *younger* for high values of $q_0$, to get the same $z = 0$ model-galaxy we would have to use shorter SFR e-folding times.

The optical depth of the models has been assumed to depend on the present-day luminosity to account for the fact that less luminous galaxies seem to have less reddening (dust) than bright galaxies (Wang 1991). The dependence on the luminosity does not make a strong effect in the predictions because the counts are mainly dominated by $L_*$ galaxies. On the other hand the extinction law is very important for the model predictions. In fact in the extreme case that the optical depth did not depend on the wavelength (grey extinction law) the dust would not produce any effect at all. To illustrate this we plot in Figures 8 and 9 the number counts and redshift distribution predictions (for the S-model) from three different dust models: $\lambda^{-2} - \tau \propto L^{0.5}$, $\lambda^{-1} - \tau \propto L^{0.5}$ and $\lambda^{-2} - \tau \propto L^0$ (i.e. no dependence on the luminosity). The two models with a $\lambda^{-2}$ extinction law give very similar predictions, as it is expected. The predictions from the model with a $\lambda^{-1}$ extinction law are *half-way* from the PLE model and the $\lambda^{-2}$ PLE+D model. The reason is that the brightening of the galaxies at high-z is now less *dimmed* by the extinction due to the weaker dependence on the wavelength.

### 3.2  K-band Counts and Colours

One of the problems of PLE models was found in the colour distribution of faint galaxies, which are predicted to be on average *bluer* than observed (although see Shanks et al. 1995). The reddening of the light produced by the dust can help to solve the problem, as shown in Figure 10. The models do not give a *perfect* fit to the data, as the predicted range of colours is wider than observed. However the observed and predicted distributions of colours are now well centered. We suggest that the discrepancies between data and model predictions might be due to the simplicity of the dust model proposed, or even to the luminosity evolution models (eg. the colours are very sensitive to changes in the IMF or metallicity).

The number counts in the K band (data from Mobasher et al. 1986; Glazebrook et al. 1993; Jenkins & Reid 1991; Gardner et al. 1993) are plotted in Figure 11, together with predictions from the $q_0 = 0.05$ and $q_0 = 0.5$ PLE+D models for comparison. To compute the counts in the K band we used the colours provided by the galaxy models for different redshifts, using afterwards the B-band luminosity function. The models agree with the observations in the magnitude range $K \sim 15 - 19$. Fainter than $K \sim 19$ the $q_0 = 0.05$ model slightly over-predicts the counts and brighter than $K \sim 15$ the observations show a steeper slope (this also happens in the B-band brighter than $B = 17.5$ using our *high* normalization). The $q_0 = 0.5$ model provides a better fit to the data, except at the faint end ($K \sim 21$) where under-predicts counts. In any case we suggest that the models fit the data reasonably, and regard the small discrepancies as due to the uncertainties in the spectrophotometric models when computing the $B - K$ colour of the model galaxies. Finally, it is worth mentioning that the inclusion of dust in modelling the counts has little effect for the K-band predictions (the ratio of optical depths between the B and K band is $\sim 1/11$).

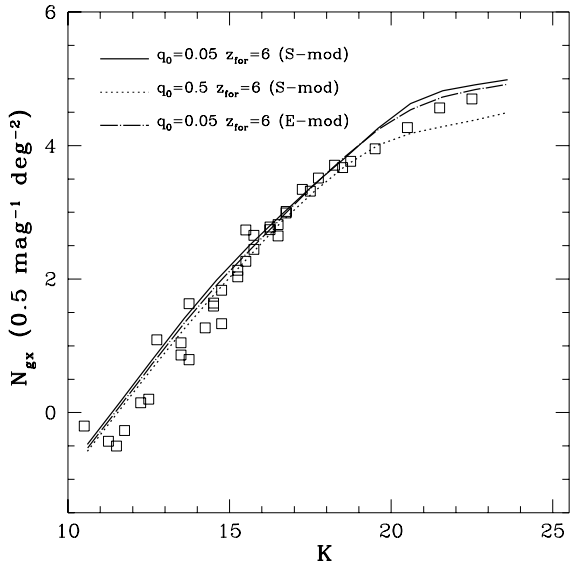

**Figure 11.** The number counts in the K band (data taken from the literature, see text for references) and predictions from PLE+D models

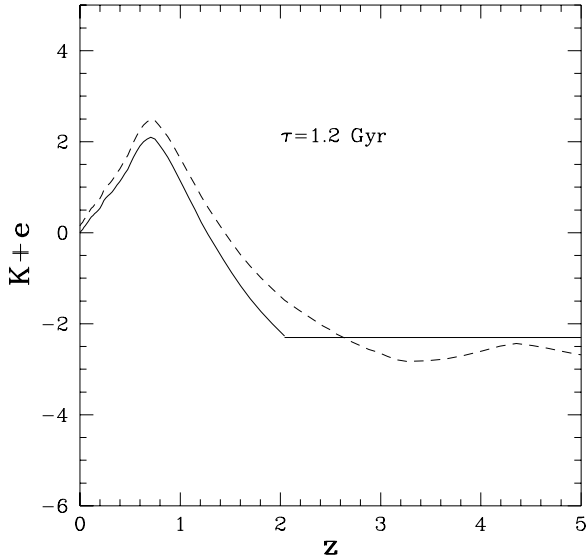

**Figure 12.** The reddened K+e corrections for the $\tau = 1.2$ Gyr model and the $zcut$ unreddened ones used by Metcalfe et al.

### 3.3 Comparison between Old PLE and New PLE+D Models

The PLE+D E-model gives similar predictions to the PLE model proposed by Metcalfe et al. (1991; 1995). As already mentioned, in this model the *excess* of counts over the predictions from non-evolving models is mainly addressed in terms of the luminosity evolution of elliptical galaxies ($\tau = 1.2$ Gyr). In Figure 12 we plot the (unreddened) $K+e$ corrections for a $\tau = 1.2$ Gyr model-galaxy, which is considered constant *(zcut)* beyond $z = 2$ as in Metcalfe et al., together with the reddened $K+e$ corrections. As can be seen, the $zcut$ and reddened $K+e$ corrections are not too different. Therefore the predictions from the two models are expected to be quite similar, as it actually happens. The $zcut$ approximation in Metcalfe et al. has then turned out to *mimic* the effect of dust in the B-band luminosity evolution.

## 4 COULD WE FIT THE COUNTS WITH A "CLOSED" MODEL?

We have seen in the previous section that simple $q_0 = 0.05$ $z_{for} = 6$ models based on the observed local LF and luminosity evolution *plus* a dust component can provide a good fit to the counts in the B band in a range of $\sim 10$ magnitudes ($B \sim 17-27$). The counts predicted in the K-band still show some small discrepancies with the data, but these are most probably due to the simplicity of the models. The models are also able to give a very good match to the observed $N(z)$, provided that the unidentified faint galaxies were at high redshifts, as suggested by Campos et al. (1995) (but also to be in agreement with the observed amplitude of $\omega(\theta)$ at faint magnitudes, Roche et al. 1993; 1995). The S-model provides a better fit to the $B : 23 - 24$ $N(z)$ distribution than the E-model, although we do not discard the E-model on this basis because we are not including dust evolution. In a *closed* Universe these models do not work, as can be seen in Figure 5. There is an *excess* of galaxies fainter than $B \sim 25$ over the predictions from the models.

A possibility to fit the faint counts with a $q_0 = 0.5$ model is by steepening the LF slope locally. In fact there is still debate about how numerous is the population of dwarf galaxies in the Universe and how this population is contributing to the counts (Cowie et al. 1992; Driver et al. 1994, 1995). Certainly we know that in local clusters of galaxies the percentage of dwarfs is as large as $\sim 30-50\%$ (Binggeli, Sandage & Tamman 1985; Phillips et al. 1987; Davies et al. 1988). On the other hand the measurements of the faint end slope of the LF are still rather uncertain.

As clearly illustrated in Driver et al. (1994), the bright galaxies start to notice the *cosmological volume effect*, i.e. the volume decreases, especially in a closed geometry, while the contribution from faint galaxies, situated at lower-z, still increases with a Euclidean slope. Then, at faint apparent magnitudes the counts start to be dominated by the intrinsically faint galaxies. However the steepening of the local LF slope does not help to solve the problem of the faint counts, because even if these models can provide nice fits to $N(m)$ they largely over-predict the number of low-z galaxies to be observed in the faint redshift surveys.

An alternative approach to solve the problem was suggested by Metcalfe et al. (1995), who changed the slope of the LF discontinuously at $z = 1$, assuming that beyond it $\alpha \sim -1.8$. As commented by these authors, this *ad hoc* solution for the closed model resembles the luminosity-dependent evolution first invoked by Broadhurst et al. (1988), but applied at much fainter magnitudes.

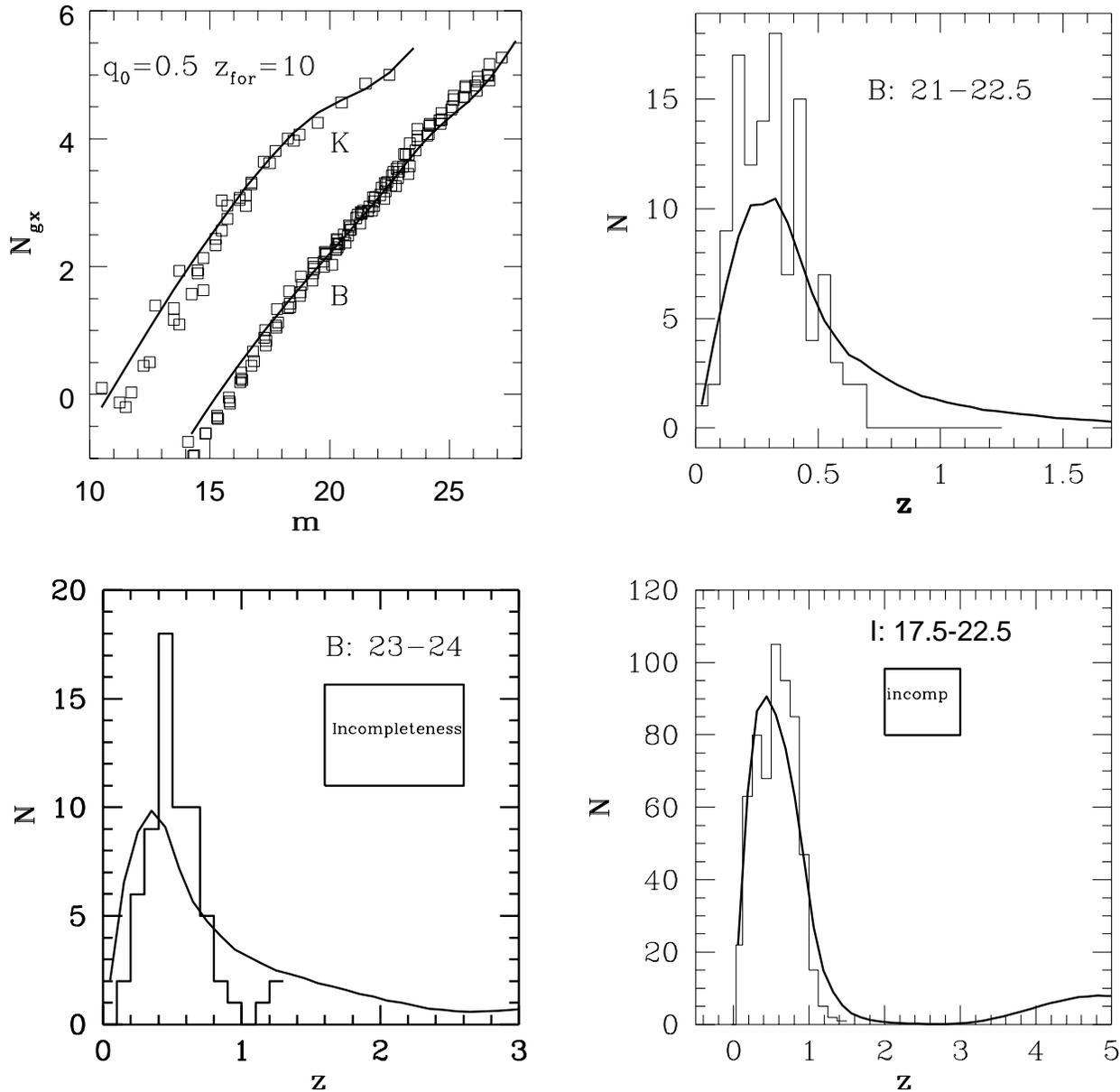

**Figure 13.** The number counts in the B and K bands, and redshift distribution for galaxies in the ranges B:21-22.5, B:23-24 and I:17.5-22.5, and predictions from a $q = 0.5$ LE+D model ($\tau_E = 0.3$ Gyr). The slope of the LF for late type galaxies is assumed to evolve with redshift as $\alpha \propto f * z$ with $f = 0.5$.

We have computed the predictions from a PLE+D S-model with $q_0 = 0.5$ and $z_{for} = 10$, although now we consider $\tau_E = 0.3$ Gyr. The B-counts are normalized at $B \sim 18$, what corresponds to a somewhat higher overall normalization, i.e. $\phi_* = 1.7 \times 10^{-2}$. In order to account for the *excess* of galaxies at faint magnitudes, we assume that the slope of the LF *only for the late type galaxies* evolves as $\alpha \propto f * z$ with $f = 0.5$ (i.e. becomes steeper at high-z; The value of $f$ has been chosen to fit the B-band counts). With this *arbitrary* evolution of the faint end of the LF we try to model (at least phenomenologically) the existence of a population of *fading dwarfs* which becomes increasingly important at higher redshifts. This model is shown in Figure 13.

As a check to the model we plot in the same Figure predictions for the K-band counts, and $N(z)$ for B- and I-

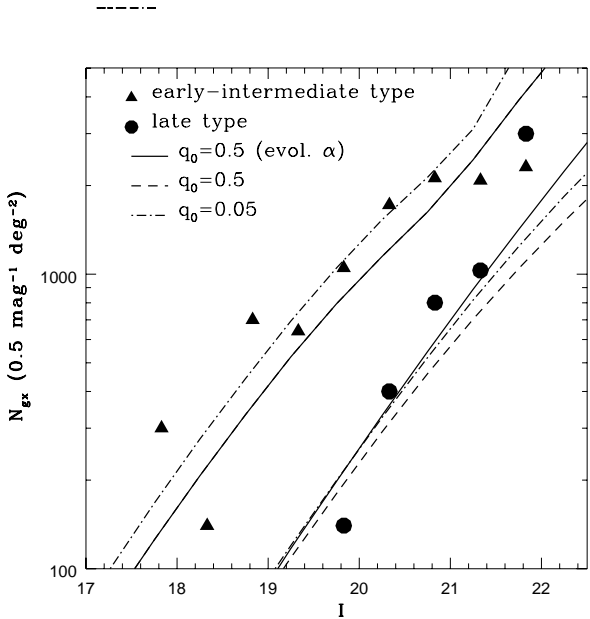

**Figure 14.** The I-band counts of Glazebrook et al. (1995b) from the HST Medium Deep Survey, for ellipticals plus spirals (triangles) and irregulars (circles). We also show predictions from our $q_0 = 0.5$ model with evolving faint-end slope (solid line), and predictions from the $q_0 = 0.05$ (dot-dashed line) and 0.5 ($z_{f}or = 6$; dashed line) shown in Figures 6 and 7

selected galaxies. The fit to the K-band counts is very good, as can be seen in the Figure. With respect to $N(z)$, the predictions are not very different to those that are provided by models *without* evolution of the LF slope, the reason being that this evolution stars to be noticiable *beyond* $B \sim 24-25$. Therefore, the predictions for $N(z)$ are in nice agreement with the data.

As a final test to the model we show in Figure 14 the I-band counts of Glazebrook et al. (1995b) from the HST Medium Deep Survey, for ellipticals plus spirals (triangles; early-intermediate types in the models) and irregulars (circles; late types) together with the model predictions. The model fits the data reasonably well except for the deepest bin ($I = 21.5 - 22$), as it under-predicts the percentage or late type galaxies while over-predicting the number of early ones. As the discrepancy occurs only at the very faint limits, we suggest (see also Shanks et al. 1995) that some of the faint galaxies identified as irregulars could rather be the progenitors of the present-day ellipticals. In the same Figure we also show predictions from the $q_0 = 0.05$ and 0.5 models shown in Figures 6, 7, 8 and 9 (S-model), except for the normalization, which is the same than for the $\alpha$-evol $q_0 = 0.5$ model for comparison. The *closed* model underpredicts the counts of late type galaxies. The *open* model fits reasonably well the data, except at the very faint limits where again the number of irregulars is under-predicted and the number of early-intermediate over-predicted. Also, the fit to the data seems to suggest the need for high-normalization (see Shanks 1990; Metcalfe et al. 1991; Shanks et al. 1995).

## 5  DISCUSSION

In order to build a realistic model to predict the counts, it is neccesary to consider the different ingredients that might have a non-negligible role in the predictions. The local LF is reasonably well known; it would seem likely that any low surface brightness population missed at bright magnitudes would also be missed at faint magnitudes, with little resulting effect on the count interpretation (but see McGaugh,1994). However, the dependence of the LF on galaxy type is still uncertain. The parameters we have used here from Metcalfe et al 1991 we believe are the best available but still require further tests in bigger galaxy redshift surveys.

With respect to the luminosity evolution of the galaxy population, over the years there has been various improvements to the models of galaxy SFR (cf. Bruzual 1983 with Bruzual & Charlot 1993). While the changes, for example, to the evolutionary tracks for the exponentially decreasing SFR's thought appropriate to early-types seem to be important for the age-colour relation for elliptical galaxies, the changes to the models themselves are less important for the interpretation of counts. However, the new availability of more detailed tracks to higher redshifts has shown problems with the implementation of eg Metcalfe et al 1991 of the original Bruzual model. The redshift cut-off applied by these authors turns out to have been important in obtaining fits to the counts. Otherwise models with exponentially increasing rates of star-formation can produce too many high redshift galaxies. Since the fits to the counts and $N(z)$ distributions are impressive with these cut-off models, one possibility we have explored here is whether the cut-off could be physically motivated by the presence of dust in galaxies cutting down the luminosity of high redshift galaxies, particularly in the bluer wavebands.

A further change to the luminosity evolution models is that the models now recommended by Bruzual and Charlot for the intermediate spiral types Sbc, Scd have exponential SFR timescales of $\sim 7$ Gyr with a single IMF Salpeter exponent. These models provide increased B luminosity evolution over those recommended by Bruzual (1983) for these spiral types; these were essentially constant SFR models with a luminosity dependent IMF slope. These models were much closer to the Sbc K-correction than the 7 Gyr model. The original reason given by Bruzual (1981) for rejecting the exponential 7Gyr model for spirals was that this model predicted too high a UV luminosity for present-day spirals. Bruzual & Charlot (1993) only discuss the fit of these models to the present day optical spectral energy distributions. Clearly it depends on how seriously the bad fit of these models in the UV is taken as to whether these models are acceptable or not. It should also be noted that the inclusion of dust may redden the spectra enough to overcome this previous objection.

The existence of these spiral models therefore leads to an alternative 'PLE' basis for explaining the B number counts, where the evolution is based on spiral together with early-type evolution. In this case early-type models can be used which have tau=0.5 Gyr which predicts less evolution for early-types at low redshifts and where the count evolution at B$\approx$ 25 is more due to the spirals. However, even the exponential tau=0.5 models, if left uncut in redshift, still overpredict the counts at B=25. Again dust may have to be invoked in these cases if an exponential model is to be used for the early-types. We have therefore investigated the effects of including a simple dust model in both the context of the original early type evolution model (E-model) and in the context of the newer spiral based evolution model (S-model).

We now summarise the results of introducing dust into both these evolutionary models for galaxy counts. In §3 we showed that even small amounts of dust can have an important effect on count models. In particular, it modifies the strength of observed luminosity evolution as higher-z galaxies are more obscured due to the combination of the extinction law plus spectral redshifting. Although our dust model we have used is very simple, we have shown how the discrepancies between both the elliptical and spiral LE models and $N(m)/N(z)$ observations are reduced when dust reddening is considered. Therefore we suggest that it is possible to fit the data with *simple luminosity evolution* models without the necessity of strong number density evolution. There could be some merging in the relatively recent past, but its rate may not be too high at $z < 1$ as we do not see much evidence of recent merging in most of the normal nearby spiral galaxies. Besides, the *extreme* merging models like those proposed by Broadhurst et al. (1992) fail to account for the scaling of the amplitude of $\omega(\theta)$ at $B < 25$, for the global population and as a function of colour (Roche et al 1993,1995).

The conclusion that PLE models with dust may fit the galaxy count data is similar to that reached by Gronwall & Koo (1995) but by a different route. Whereas these authors assume that the local LF is not well defined, keeping it as a free parameter, we take as a fixed parameter the observed LF for the three types of galaxies divided according to their colours (Metcalfe et al. 1991).

The present spiral/dust model may give a better fit to the $N(z)$ data than the elliptical/dust model at least for lower levels of the galaxy count normalisation. However, higher dust levels at higher redshift may allow the early-type models to get around this problem. Also as argued by Shanks 1990, a higher normalisation for the counts may reduce the high redshift component predicted by the early-type model to more acceptable limits. Shanks et al 1995 also argue that the B-K colour distributions at K=20-23 tentatively favour strong B band evolution for early-type galaxies, which may not be currently predicted by the spiral/dust model. As the deep B and K data improve this could prove a crucial discriminatory test between these two PLE models.

Despite the apparent importance of dust for the galaxy count models, its modelling is rather difficult as little is known, for example, about the evolution of dust in galaxies. Here we have considered a very simple non-evolving model, which is the same for *all* galaxies. This is clearly a very rough approximation, as in particular we know that the amount of dust in present-day ellipticals is rather small. On the other hand it is for the galaxies at high-z where the dust becomes increasingly important due to the redshifting of the spectral light. Finally, not only the amount of dust but the extinction law could change from galaxy to galaxy, or even with time. Questions like these about the dust content of evolving galaxies clearly require further investigation.

Although the dust improves the fit of both the spiral and elliptical models at $B < 25$, at fainter magnitudes it leaves the question of the poor fitting of the $q_0 = 0.5$ model relatively unaffected. Such models still require evolution in the slope of the luminosity function at faint limits if they are to fit the $B > 25$ counts, as previously noted by Metcalfe et al. (1995). All that we have shown is that it is possible to find model parameters including dust which fit the faintest count data in photometric bands ranging from B to K. In particular we note the good fit to the K-band data for $K \sim 15 - 23$. The physical interpretation of the steepening of the LF slope at $B > 25$ and $z > 1$ may correspond to an era of demerging of faint galaxies. The observations of Cowie of "chain" galaxies at $z > 1$ may support this idea that the era of galaxy merging may have been at these higher redshifts. Alternatively it could suggest that at high redshifts dwarf galaxies undergo more rapid luminosity evolution than giant galaxies. The small mass of dwarf galaxies might not be able to prevent their gas being removed in a galactic wind following the overlap of supernova explosions and this could dim these galaxies preferentially. (Dekel & Silk 1986). These types of non-PLE models are similar to those discussed by Broadhurst et al 1988 but applied at $B > 25$ rather than B=21 where we now believe there may be no need for these more sophisticated models.

We finally note that the low $q_0$ model continues to give a reasonable fit to the counts even at these faint magnitudes without the requirement of the steepening of the luminosity function slope.

## 6 CONCLUSIONS

(i) Luminosity evolution models with an exponentially decreasing SFR for early type galaxies require a redshift cutoff which may mimic the effect of dust.

(ii) New models from Bruzual & Charlot (1993) allow stronger evolution for spirals than previously and this may make it possible to construct luminosity evolution models based on spirals as well as early type galaxies which can fit number count data.

(iii) The inclusion of dust improves the possibility for both the spiral and the early-type PLE models to fit $N(m)$ and $N(z)$ data to B=25, with no need to invoke merging or luminosity dependent evolution to this magnitude limit.

(iv) At fainter magnitudes ($B > 25$), the $q_0 = 0.5$ model continues to require a steepening of the luminosity function at high redshift to obtain a reasonable fit to the counts.

### Acknowledgments

AC is funded by an EC Research Fellowship. We acknowledge useful discussion with N. Metcalfe, R. Fong, R. de Jong and J. Gardner.

## REFERENCES

Arimoto, N. & Yoshii, Y. 1986, A& A 164, 260.
Arimoto, N. & Yoshii, Y. 1987, A& A 173, 23.
Babul, A. & Rees, M.J. 1992, MNRAS 255, 346.
Binggeli, B., Sandage, A. & Tamman, E.A. 1985, AJ 90, 1681.
Broadhurst, T., Ellis, R.S. & Glazebrook, K. 1992, Nature 355,55.

Bruzual, B. 1981, PhD thesis, Univ. of California at Berkeley.
Bruzual, G. 1983, ApJ 273, 105.
Bruzual, G. & Charlot, S. 1993, ApJ 405, 538.
Bruzual, G. & Kron, R.G. 1980, ApJ 241, 25.
Campos, A., Shanks, T., Metcalfe, N., Roche, N. & Tanvir, N.R. 1995, MNRAS, submitted.
Crampton, D., Le Fevre, O., Lilly, S.J. & Hammer, F. 1995, preprint astro-ph/9507014.
Colless, M., Ellis, R.S., Taylor, K. & Hook, R.N. 1990 MNRAS 244, 408.
Colless, M., Ellis, R.S., Taylor, K. & Peterson, B.A. 1993 MNRAS 261, 19.
Couch, W.J. & Newell, E.B. 1984, ApJ Suppl 56, 143.
Cowie, L.L., Songalia, A. & Hu, E.M. 1992, Nature 354, 460.
Cowie, L.L., Hu, E.M. & Songalia, A. 1995, Nature submitted.
Davies, J.I., Phillips, S., Cawson, M.J., Disney, M.J. & Kibblewhite, E. 1988, MNRAS 232, 239.
Dekel, A. & Silk, J. 1986, ApJ 303, 39.
Draine, B.T. & Lee, H.M. 1984, ApJ 285, 89.
Driver, S.P., Phillips, S., Davies, J.I., Morgan, I. & Disney, M.J. 1994, MNRAS 266, 155.
Driver, S.P., Windhorst, R.A., Ostrander, E.J., Keel, W.C., Griffiths, R.E. & Ratnatunga, K.U. 1995, ApJ 449, 23.
Gardner, J.P., Cowie, L.L. & Wainscoat, R.J. 1993, ApJ 415, L9.
Gardner, J.P. 1995, ApJ submitted.
Glazebrook, K., Ellis, R., Colless, M., Broadhurst, T., Allington-Smith, J., Tanvir, N.R. & Taylor, K. 1995, MNRAS 273, 157.
Glazebrook, K., Ellis, R., Santiago, B. & Griffiths, R. 1995, MNRAS in press.
Guiderdoni, B. & Rocca-Volmerange, B. 1990, A&A 227, 362.
Guiderdoni, B. & Rocca-Volmerange, B. 1991, A&A 252, 435.
Gronwall, C. & Koo D.C. 1995, ApJ 440, L1.
Infante, L., Pritchet, C. & Quintana, H. 1986, AJ 91, 217.
Jarvis, J.F. & Tyson, J.A. 1981, AJ 86, 476.
Jenkins, C.R. & Reid, I.N. 1991, AJ 101, 1595.
Jones, L.R., Fong, R., Shanks, T., Ellis, R.S. & Peterson, B.A. 1991, MNRAS 249, 481.
King, C.R. & Ellis, R.S. 1985, ApJ 288, 456.
Koo, D.C. 1981, PhD thesis, Univ. of California at Berkeley.
Koo, D.C. 1986, ApJ 311, 651.
Kron, R.G. 1978, PhD thesis, Univ. California at Berkeley.
Lilly, S., Cowie, L.L. & Gardner, J.P. 1991, ApJ 369, 79.
McGaugh, S. 1994, Nature 367, 538.
Metcalfe, N., Shanks, T., Fong, R. & Jones L.R. 1991, MNRAS 249, 498.
Metcalfe, N., Shanks, T.& Fong, R. 1995 MNRAS in press.
Mobasher, B., Ellis, R.S. & Sharples, R.M. 1986, MNRAS 223, 11.
Murante, G., Yepes, G., Klypin, A.A. & Campos, A. 1995, in preparation.
Neuschaefer, L., Ratnatunga, K.U., Griffiths, R.E. & Valdes, F. 1995, PASP 107, 590.
Phillips, S., Disney, M.J., Kibblewhite, E.J. & Cawson, M.G.M. 1987, MNRAS 229, 505.
Roche, N., Shanks, T., Metcalfe, N. & Fong, R. 1994 MNRAS 263, 360.
Roche, N., Shanks, T., Metcalfe, N. & Fong, R. 1995, in preparation.
Shanks, T., Stevenson, P.R.F., Fong, R. & MacGillivray, H.T. 1984, MNRAS 206, 767.
Shanks, T. 1990. In "The Galactic and Extragalactic Background Radiation", p. 269, eds. S. Bowyer and C. Leinert. Dordrecht: Kluwer.
Shanks, T., Campos, A., Metcalfe, N. & Fong, R. 1995, in preparation.
Tinsley, B.M. 1972, A&A 20, 383.
Tinsley, B.M. 1980, ApJ 241, 41.
Tyson, J.A. 1988, AJ 96, 1.
Wang, B. 1991, ApJ 383, L37.
Wyse, R.F.G. 1985, ApJ 299, 593.
Yoshii, Y. & Takahara, F. 1988, ApJ 326, 1.